\documentclass{llncs}
\usepackage{latexsym}
\usepackage{graphicx}
\usepackage{bsymb}
\usepackage{varioref}
\usepackage{amssymb}
\usepackage{amsmath}
\usepackage{amsbsy}
\usepackage{moreverb,color}
\usepackage{multirow}
\usepackage{subfigure}

\usepackage{listings} 
\usepackage{xcolor} 
\usepackage{picinpar}

\newcommand{\Comment}[1]{}

 \newcounter{ReqEnv}

\newcounter{ReqFun}

\title{Formal Proof of the Weak Goodstein Theorem} 

\author{Jean-Raymond Abrial}
\institute{ Marseille, France}

\begin{document}

\maketitle

\section{Motivation}
 For many years, I have been interested in introducing students to the development of complex systems by means of modelling and refinement. To this end, I did not find anything better than presenting {\em many examples of system developments}.  This is due to my inability to propose a unified theoretical treatment on this matter.
 
Of course, in these examples, I am always pointing out the importance of using some systematic mathematical approaches. However, I figured out that my examples were not explicit enough on how (mechanical) proofs are performed. So, besides courses presenting these examples and also some courses in various forms of proofs (propositional calculus, first order predicate calculus, set theory), I decided to study the {\em work of professional mathematicians}, thinking that it could be good examples for students. 

I must say that I was a bit disappointed by what I discovered: proofs made by mathematicians, as presented in textbooks, are sometimes (for me) difficult to follow in details and thus could have some bad effects on students. As a consequence, I decided to reconstruct by myself some of the interesting proofs I found in the mathematical literature. 

Among the works I already studied and reconstructed are the theorem of Zermelo, the theorem of Cantor-Bernstein, the planar graph theorem of Kuratowski, the topological proof of the infinity of primes of F\"urstenberg, the intermediate value theorem of Bolzano, the Archimedean property of the set of Real numbers, and others. 

More recently, I found that the Goodstein theorem was also very interesting. The purpose of this short note is to give some information about this theorem and the way I introduce a weak form of it to students.

\section{The Goodstein Theorem}

The theorem stated and proved in 1944 by Goodstein \cite{gd}, is quite counterintuitive. To explain why, let me consider a weak form of it called, for this
reason, the {\em weak Goodstein theorem}.

Given a number written in base 2 such as 25, that is 11001 ($2^4 + 2^3 + 1$), we
transform it by considering the same notation but this time in base 3, that is
$3^4 +3^3 +1 = 109$. We then subtract 1, yielding 108. We write now this number in
base 4, and subtract 1 again, yielding 319. With base 5, we obtain 717, with base
6, 1423. We continue like this: increasing the base and decreasing the result. As
can be seen from what is already mentioned, the successive numbers obtained in
this way seem to grow up very rapidly: 25, 108, 319, 717, 1423, . . . . Nevertheless,
the theorem says that this sequence eventually {\em decreases and terminates at 0}.

The {\em strong Goodstein theorem} is a little more general than the weak form what we have just
described in the previous paragraph, and it is even more counterintuitive. It
is not expressed with the classical base notation, as was the weak Goodstein
theorem, but rather with the, so-called, {\em hereditary base notation} (explained in
section \ref{he}).

Proofs of these theorems in the literature \cite{gd} \cite{rt} \cite{cd} \cite{gs} \cite{ki} make use of transfinite ordinal
numbers. I found that this approach is rather complicated. So, I am looking for another (simpler) possibility. So far, I partially fail, at least for the strong Goodstein theorem. Its weak form however can be proved in a simple fashion. This is what I present here.

\section{Hereditary Base Notation} \label{he}

By using the classical notation in base $n$ (where $n$ is a natural number greater
than 1), any natural number $a$ is written as follows (base\_n($a$)):
\[
\hbox{base\_n}(a) = a_l .n^l + \cdots  + a_i .n^i + \cdots + a_0 .n^0
\]\noindent where all $a_i$ are natural numbers smaller than $n$. As a simplification for this
written form, we omit $0.n^i$, we write $1.n^i$ as $n^i$ , $n^1$ as $n$, and $n^0$ as 1. As an
example, 25 (that is 16+8+1) is written as follows in base 2:
\[
\hbox{base\_2}(25) = 2^4 + 2^3 + 1
\]
By using a notation in hereditary base $n$,
the exponents $i$ used in the notation in base $n$ are also written in base $n$ and so
on ($\hbox{h\_base\_n}(a)$):
\[
\hbox{h\_base\_n}(a) = a_l .n^{\hbox{h\_base\_n}(l)} + \cdots + a_i .n^{\hbox{h\_base\_n}(i)} + \cdots + a_0 .n^0
\]
So, all natural numbers appearing when using the hereditary base $n$ notation are smaller than or equal to $n$. As an example, 25 is written
as follows in hereditary base 2:
\[
\hbox{h\_base\_2}(25) = 2^{2^2} + 2^{2+1} + 1
\]
\section{Data Structures for Base Notations} \label{ds}
As we all know, writing a natural number in a certain base consists quite often
in removing the base when it is obvious. As a result, we have just a sequence of
digits (all smaller than the base). For example, 25 in base 2 is simply written:
11001. Such a sequence is organised as follows: it starts at index 0 and goes
from right to left. Each number at index $i$ corresponds to the factor used with
exponent $i$. This is illustrated in Fig.1.
\begin{center}
 \includegraphics[scale=.31]{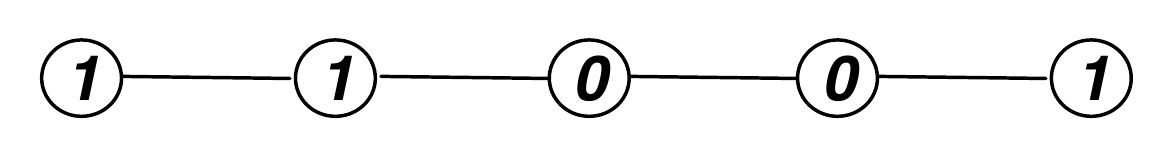} \\ \ \\
Fig. 1. Sequence representation of $25 = 1.2^4 + 1.2^3 + 0.2^2 + 0.2^1 + 1.2^0$
\end{center}
Notice that in this representation we do not omit components of the forms $0.2^i$,
nor do we omit the 1 in those components of the form $1.2^i$. We notice that this
sequence representation does not depend on the base: $11001_{10}$ corresponds to the
same sequence (but not the same number) as $11001_2$ or $11001_3$. 

We wonder whether we could do the same (omitting the base) when using the
hereditary technique. How could we remove all occurrences of 2 in 25 written in
hereditary base 2: $2^{2^2} + 2^{2+1} + 1$? Here is a more elaborate example: $774840988_{10}$
in base 3 is $2.3^{18} + 3^2 + 2$ and in hereditary base 3 it is $2.3^{2.3^2} + 3^2 + 1$. {\em How
could we remove all occurrences of 3?}

The idea is to observe that we have three operations in such a representation: addition, exponentiation and multiplication by a factor. The outcome is
a {\em binary tree}, where the horizontal branch corresponds to addition, the vertical
branch to exponentiation, and finally the multiplication by a certain factor is
just indicated by writing this factor in the corresponding node of the tree. The
tree representation of $2.3^{2.3^2} + 3^2 + 1$ (in hereditary base 3) is shown in Fig. 2.
\begin{center}
 \includegraphics[scale=.31]{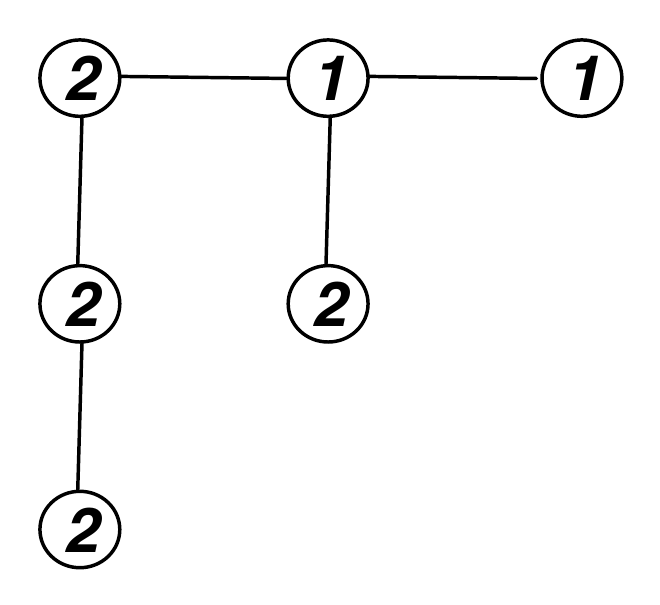} \\
Fig. 2. Tree representation of $2.3^{2.3^2} + 3^2 + 1$ in hereditary base 3
\end{center}
As was the case in the previous section for the sequence representation, it is also
very important to notice that the tree representation introduced in this section does not depend on the
hereditary base. For example, the number $2.4^{2.4^2} + 4^2 + 1$ is represented by the
same tree in hereditary base 4 as is $2.3^{2.3^2} + 3^2 + 1$ in hereditary base 3. To
make the distinction between the two, it is necessary to write next to the tree
the hereditary base that is used.

In coming sections, I will use the sequence data structure in order to prove the weak Goodstein theorem. I was hoping to use the tree data structure to prove the strong Goodstein theorem. But, so far, I failed.

\section{Decreasing Sequence of Natural Numbers} \label{dc}

Before engaging in a study of the weak Goodstein theorem in the next section, it is
worth considering a simple decreasing sequence of natural numbers. The purpose of this highly simplified case is to show the main mechanism at work, namely lexicographical ordering. It is based
on the following simple lemma valid for all positive natural numbers $x$ and $n$:
\begin{equation}
x^n - 1 = (x -1).x^{n-1} + (x-1).x^{n-2} + . . . + (x-1).x^1 + (x-1).x^0) \tag{\bf Lemma 1}
\end{equation}
This lemma is easily provable by induction on $n$. Applying this lemma to 
decreasing $24 = 2^4 + 2^3$, we obtain the following:
\[
\hbox{Decreasing}(2^4 + 2^3 ) = 2^4 + 2^3 - 1 = 2^4 + (2^3 - 1) = 2^4 + 2^2 + 2 + 1
\]
This is illustrated in Fig. 3.
\begin{center}
 \includegraphics[scale=0.31]{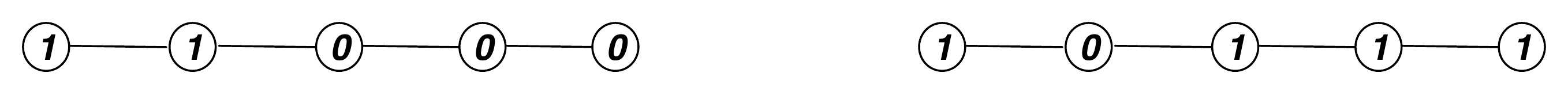} \\ \ \\
Fig. 3. Sequence representations of $2^4 + 2^3$ and $2^4 + 2^3 - 1$
\end{center}
Likewise, in base 3 we have:
\[
\hbox{Decreasing}(3^4 + 3^3 ) = 3^4 + 3^3 - 1 = 3^4 + (3^3 - 1) = 3^4 + 2.3^2 + 2.3 + 2
\]
This is illustrated in Fig. 4.
\begin{center}
 \includegraphics[scale=.31]{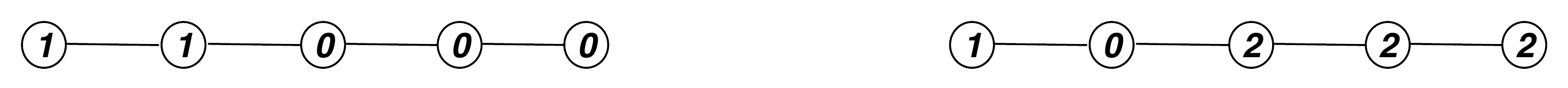} \\ \ \\
Fig. 4. Sequence representations of $3^4 + 3^3$ and $3^4 + 3^3 - 1$
\end{center}
It is interesting to observe the difference between $2^4 + 2^3 - 1$ in Fig. 3 and $3^4 + 3^3 - 1$ in Fig. 4.
They both come from the same sequence, either understood to be in base 2 or in base
3. In the second one, all 1 used in the first one are replaced by 2. This is because
2 - 1 = 1 and 3 - 1 = 2.

By decreasing successively an initial number written by means of some base,
we obtain a certain sequence and we can prove that such a sequence ends up
with the natural number 0. More precisely, we have a {\em lexicographical ordering}.

\section{Informal Proof of the Weak Goodstein Theorem} \label{wg}

In the case of weak Goodstein sequences, decreasing is done in the same way as in the
previous section except that we increase the base before decreasing. Applying
this result to $2^4 + 2^3$, we obtain the following:
\[
\hbox{Weak Goodstein Decreasing}(2^4 + 2^3) = 3^4 + (3^3 -1) = 3^4 + 2.3^2 + 2.3 + 2
\]
Although  a weak Goodstein sequence seems to be increasing
very rapidly, it happens that such a sequence obtained by applying this process in
turn ends up eventually at 0. As a matter of fact, we have the following theorem:
\begin{center}
\em Any weak Goodstein sequence eventually terminates at 0
\end{center}
{\bf Informal Proof}: We already know that increasing the base does not modify the representation of a number as a sequence. Then decreasing by one after increasing the
base just makes the resulting sequence {\em lexicographically smaller} than the previous one, hence the final result.\\

\noindent Here is the beginning of the weak Goodstein sequence starting at $1000_2$:
\[
\begin{array}{l}
1000_2 \ \ 222_3 \ \ 221_4 \ \ 220_5 \ \ 215_6 \ldots 210_{11} \ \ 20(11)_{12}  \ldots 200_{23} \ \ 1(23)(23)_{24}  \ldots \\
1(23)0_{47} \ \ 1(22)(47)_{48}  \ldots1(22)0_{95} \ \ 1(21)(95)_{96}  \ldots 1(21)0_{191} \ \ 1(20)(191)_{192} \\
\ldots 1(20)0_{383} \ \ 1(19)(383)_{384}  \ldots 1(19)0_{767} \ \ 1(18)(767)_{768}  \ldots 1(18)0_{1535}  \ldots
\end{array}
\]
On this sequence, {lexicographical decreasing} can be seen independently from
the current base.

\section{A More Formal Treatment of the Weak Goodstein Theorem}
\subsection{Constructing the sequence $\text{seq}_b(n)$ associated with a number $n$ in base $b$}
\[
\begin{array}{l}
\text{seq}_b(n) = \left \{ \begin{array}{l}
                                       \text{seq}_b(n  \text{\;div\;}  b) \leftarrow n \text{\;mod\;}b \ \ \ \ \ \ \hfill{\text{if}\ \ n \geq b} \\
                                       n \hspace{14.3em} \hfil{\text{if} \ \ n<b}
                                       \end{array} \right.
\end{array}
\]
\[
\begin{array}{ll}
\text{seq}_2(25) & = \;\text{seq}_2(12) \leftarrow 1\\
& = \;\text{seq}_2(6) \leftarrow 0  \leftarrow 1\\ 
& = \;\text{seq}_2(3) \leftarrow 0 \leftarrow 0  \leftarrow 1\\ 
& = \;\text{seq}_2(1) \leftarrow 1 \leftarrow 0 \leftarrow 0  \leftarrow 1\\ 
& = \;1  \leftarrow {1} \leftarrow {0} \leftarrow {0}  \leftarrow {1}\\ 
\end{array}
\] 
\subsection{Value $\text{val}_b(s)$ of the number associated with a sequence $s$ in base $b$}
\[
\begin{array}{l}
\text{val}_b(s \leftarrow n)  =  b.\text{val}_b(s) + n  \hspace{3em} \text{where} \;\; n<b \\ 
\text{val}_b(n)  =  n
\end{array}
\]
\[
\begin{array}{ll}
\text{val}_2({1}  \leftarrow {1} \leftarrow {0} \leftarrow {0}  \leftarrow {1}) & = 2.\text{val}_2(1  \leftarrow 1 \leftarrow 0 \leftarrow 0)+1 \\
& = 2^2.\text{val}_2(1  \leftarrow 1 \leftarrow 0)+0+1 \\ 
& = 2^3.\text{val}_2(1  \leftarrow 1)+0+0+1 \\ 
& = 2^4.\text{val}_2(1)+ 2^3+0+0+1 \\ 
& = {1}.2^4+ {1}.2^3+{0}.2^2+{0}.2^1+{1}.2^0 \\ 
& = 16 + 8 + 1 \\
& = 25
\end{array}
\]
\subsection{The Weak Goodstein loop}
Next is the loop producing successive elements of the weak Goodstein sequence in the variable $n$
.
\[
\begin{array}{|l|} \hline \\
\;\;\;n := \hbox{some natural number}; \\
\;\;\;b := 2; \\
\;\;\;{\hbox{\bf while}} \ \ n \neq 0 \ \ {\hbox{\bf do}}\\
\;\;\; \ \ \ n := \text{val}_{b+1}(\text{seq}_b(n))-1;\;\\
\;\;\; \ \ \ b \ := b+1 \\
\;\;\;{\hbox{\bf end}}
\\ \\ \hline
\end{array}
\]
Now, the question is: {\em does this loop terminate?}

\subsection{Formal Development Outline}

Here is the way I develop the course for students. As can be seen, it allows me to develop various formal techniques.\\

\noindent 1. How to prove {\em loop termination}
\\ 
2. Definitions of {\em well-founded relations} (demo with Rodin toolset)
\\ 
3. Various ways of {\em proving well-foundedness} (demo with Rodin toolset)
\\ 
4. How to prove the {\em termination of the Weak Goodstein loop}
\\
5. Refresher on {\em strong well ordering} relations (demo with Rodin toolset)
\\ 
6. Refresher of various {l\em exicographical ordering relations} (demo with Rodin toolset)
\\ 
7. {\em Final Proof} for the Weak Goodstein Loop

\end{document}